\documentclass[sigconf, screen]{acmart}
\usepackage{array}
\usepackage{xspace}
\usepackage{graphicx}
\usepackage{subcaption}
\usepackage{xcolor}

\usepackage{xcolor}
\usepackage{enumitem}

\usepackage{amsmath,amssymb}

\usepackage{changepage}
\usepackage{framed}

\makeatletter
{\par\unskip\endMakeFramed}
\makeatother
\definecolor{formalshade}{HTML}{F9E3DF}

\definecolor{darkblue}{rgb}{0.2, 0.2, 0.2}

\newenvironment{formal}{%
\vspace{-5pt}
\def\FrameCommand{%
\vspace{-10pt} \hspace{1pt}%
{\color{darkblue}\vrule width 2pt}%
{\color{formalshade}\vrule width 4pt}%
\colorbox{formalshade}%
}%
\MakeFramed{\advance\hsize-\width\FrameRestore}%
\noindent
\hspace{-1pt}
\begin{adjustwidth}{}{7pt}

}{%
\vspace{0pt}
\end{adjustwidth}\endMakeFramed%
}

\usepackage{tikz}

\newcommand{\llmsystem}{\textit{LLM-powered Codebase Understanding System}\xspace}

\newcommand{\yellow}[1]{\colorbox{orange!20}{\textbf{#1}}}
\newcommand{\lightblue}[1]{\colorbox{cyan!20}{\textbf{#1}}}
\newcommand{\red}[1]{\colorbox{purple!20}{\textbf{#1}}}

\newif\ifcomment
\commenttrue   

\ifcomment
  \newcommand{\gao}[1]{\textcolor[rgb]{0.6,0,0.2}{Jie: #1}}
  \newcommand{\cao}[1]{\textcolor[rgb]{0.6,0,0.2}{Cao: #1}}
  \newcommand{\todo}[1]{\textcolor[rgb]{0.859,0.569,0.043}{TODO: #1}}
  
  \newcommand{\delete}[1]{\textcolor[rgb]{0.8,0.8,0.8}{#1}}

\else
  \newcommand{\gao}[1]{}
  \newcommand{\cao}[1]{}
  \newcommand{\todo}[1]{}
  
  \newcommand{\delete}[1]{}
\fi

\AtBeginDocument{%
  }

\copyrightyear{2026}
\acmYear{2026}
\setcopyright{cc}
\setcctype{by-nc-nd}
\acmConference[ICPC '26]{34th IEEE/ACM International Conference on Program Comprehension}{April 12--13, 2026}{Rio de Janeiro, Brazil}
\acmBooktitle{34th IEEE/ACM International Conference on Program Comprehension (ICPC '26), April 12--13, 2026, Rio de Janeiro, Brazil}
\acmPrice{}
\acmDOI{10.1145/3794763.3794822}
\acmISBN{979-8-4007-2482-4/2026/04}

\begin{document}

\title{Understanding Codebase like a Professional! Human–AI Collaboration for Code Comprehension}

\author{Jie Gao}
\affiliation{%
  \institution{Johns Hopkins University}
  \city{Baltimore, MD}
  \country{USA}
}
\email{jgao77@jh.edu}

\author{Yue Xue}
\affiliation{%
  \institution{Metatrust Labs}
  \country{Singapore}
}
\email{xueyue@metatrust.io}

\author{Xiaofei Xie}
\affiliation{%
  \institution{Singapore Management University}
  \country{Singapore}
}
\email{xfxie@smu.edu.sg}

\author{Junming Cao}
\affiliation{%
  \institution{Fudan University}
  \country{Shanghai, China}
}
\email{21110240004@m.fudan.edu.cn}

\author{SoeMin THANT}
\affiliation{%
  \institution{Singapore-MIT Alliance for Research and Technology}
  \country{Singapore}
}
\email{soemin.thant@smart.mit.edu}

\author{Erika Lee}
\affiliation{%
  \institution{University of California, San Diego}
  \city{San Diego, CA}
  \country{USA}
}
\email{erl015@ucsd.edu}

\author{Bowen Xu}
\affiliation{%
  \institution{North Carolina State University}
  \city{Raleigh, NC}
  \country{USA}
}
\email{bxu22@ncsu.edu}

\renewcommand{\shortauthors}{Jie Gao, Yue Xue, Xiaofei Xie, Junming Cao, SoeMin THANT, Erika Lee, Bowen Xu}

\newcommand{\systemabb}{\texttt{CodeMap}\xspace}

\begin{abstract}
Understanding an unfamiliar codebase is an essential task for developers in various scenarios, such as during the onboarding process. Especially when the codebase is large and time is limited, achieving a decent level of comprehension remains challenging for both experienced and novice developers, even with the assistance of large language models (LLMs). Existing studies have shown that LLMs often fail to support users in understanding code structures or to provide user-centered, adaptive, and dynamic assistance in real-world settings.

To address this, we propose learning from the perspective of a unique role, code auditors, whose work often requires them to quickly familiarize themselves with new code projects on a weekly or even daily basis. To achieve this, we recruited and interviewed 8 code auditing practitioners to understand how they master codebase understanding. 
We identified four design opportunities for an LLM-based codebase understanding system: supporting cognitive alignment through automated codebase information extraction, decomposition, and representation, as well as reducing manual effort and conversational distraction through interaction design.

To validate these four design opportunities, we designed a system prototype, \systemabb, that provides dynamic information extraction and representation aligned with the human cognitive flow and enables interactive switching among hierarchical codebase visualizations. To evaluate the usefulness of our system, we conducted a user study with nine experienced developers and six novice developers. Our results demonstrate that \systemabb significantly improved users’ perceived intuitiveness, ease of use, and usefulness in supporting code comprehension, while reducing their reliance on reading and interpreting LLM responses by 79\% and increasing map usage time by 90\% compared to the static visualization analysis tool. It also enhances novice developers’ perceived understanding and reduces their unpurposeful exploration. The insights derived from our interviews and the design of \systemabb can inspire future LLM-based research on code comprehension, such as onboarding support systems. 
\end{abstract}




\maketitle

\section{Introduction}
\label{sec:introduction}
A decent understanding on the codebase is the foundation for developers to work productively. A study found that on average developers spend around 58\% of their time on program comprehension activities \cite{xia2017measuring}.
Especially, as software evolves, codebases are growing larger and becoming increasingly complex, gaining understanding becomes more critical to properly perform further software development and maintenance operations.
Codebase comprehension remains challenging even in the era of large language models (LLMs).
For example, Reefman et al. \cite{reefman2024using} found that LLM tools cannot provide the business logic necessary to understand a codebase. Similarly, Alebachew et al. \cite{berhanu2025ai} pointed out that LLM tools also fail to offer the adaptive, dynamic, and interactive strategies that developers use in real-world settings.

To systematically aid developers (regardless of experience level) in codebase-level understanding, we first observed that a special role, i.e., domain practitioners in code auditing (also known as code auditors), whose job is to frequently onboard new codebases (i.e., weekly or daily) to examine the codebase quality, such as inspecting potential vulnerabilities and write reports~\cite{cybersecurityguide_securitycodeauditor_2025}---possess valuable tacit knowledge that experienced and novice developers (i.e., onboarding monthly or quarterly) may lack. This tacit knowledge, with best practice, is valuable and could benefit the developers' activities related to codebase understanding.
Motivated by the above, we aim to answer the following key question:

\vspace{1mm}
\begin{formal}
\textit{What are the common strategies of professional code auditors in effective codebase-level understanding? Can we reuse them?}
\end{formal}
\vspace{2mm}

\begin{figure*}[h]
    \centering
    \includegraphics[width=\linewidth]{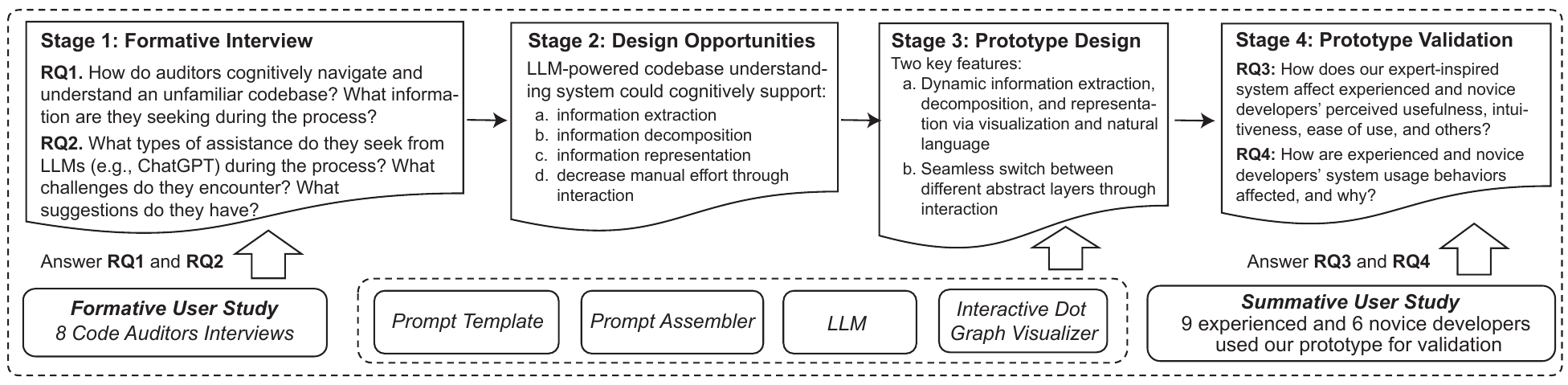}
    \vspace{-6mm}
    \caption{Methodology overview.}
    \label{fig:method}
\end{figure*}

To answer this question, we applied a four-stage method (see Section \ref{sec:method} and Figure \ref{fig:method} for details) from Formative interview $\Rightarrow$ Design opportunities development $\Rightarrow$ Prototype design $\Rightarrow$ Prototype validation, which is commonly used in human–computer interaction (HCI) research \cite{helander2014handbook, nngroup_formative_summative_2019} (see Section~\ref{sec:method} and Figure~\ref{fig:method} for details). This process enabled us to derive system design insights grounded in practitioners’ needs and practices, ensuring that the resulting system effectively satisfies user requirements. Moreover, our method uniquely focuses on code auditing practitioners, allowing us to extract practical insights from experts with extensive experience in codebase understanding. This distinguishes our approach from prior studies that rely primarily on literature \cite{berhanu2025ai} or infrequent-onboarding developers \cite{reefman2024using}. Unlike works emphasizing tool comprehensiveness and rigorous evaluation, our study prioritizes user-centered design learning from experts, using the prototype as a validation of the design insights rather than as a full-fledged product evaluation.

Through thematic analysis, we identified key patterns in how code auditors understand codebases: their understanding flows from global to local levels, and along this flow, they sequentially seek information about the project overview, codebase structure and business logic, and local functions and variables. We also observed how they used LLMs (e.g., ChatGPT) \footnote{This interview study was conducted in 2024, when ChatGPT was among the most advanced tools available for AI-assisted code reading. More recent tools such as Cursor and Trae may offer somewhat different capabilities, but the core strategies identified here remain the same.} to support this process and the challenges they encountered while doing so. Based on these findings, we derived key design opportunities: LLM systems should facilitate information extraction, decomposition, and representation that align with this hierarchical understanding flow, and support users’ interaction within it. Finally, we implemented a prototype based on identified design opportunities, and user validation with both experienced and novice developers showed that \systemabb enhanced overall user experience and increased feature engagement.

In summary, the main contributions of our work are as follows:

\begin{itemize}
    \item We empirically studied the best practice of how code auditors understand a given new codebase, what are their strategies to perform effective codebase-level understanding. We summarized the patterns of understanding flow, challenges, and expectations with detailed explanations.
    \item We translated our findings into design opportunities for \llmsystem. Building on these, we designed and implemented a prototype, \systemabb, with two key features: (i) dynamic information extraction and representation aligned with users’ understanding flow, and (ii) seamless switching between different abstraction layers through user interaction. 
    \item We empirically validated \systemabb with nine experienced and six novice developers. The results showed that \systemabb improved experienced users’ perceived intuitiveness and ease of use, reduced their reliance on reading and interpreting LLM responses by 79\%, and increased map usage time by 90\% compared with the static visualization analysis tool. For novice developers, \systemabb enhanced code comprehension and reduced unpurposeful exploration.
\end{itemize}

\section{Motivation and Research Questions}


\subsection{Why Code Auditor?}

Code Auditing is the thorough analysis of source code to identify security vulnerabilities and deviations from best practices. It can be traced back to 1976, when Michael Fagan introduced the "Fagan Inspection" \cite{fagan1976design}, which formalized a six‑step code inspection workflow that reduced defects up to 90\%. It often contains thorough codebase understanding process. Specifically, auditors must often familiarize themselves with an unfamiliar project both globally (holistic enough) and line by line (detailed enough) on a weekly or even daily basis. Through this process, auditors must develop a deep, structured grasp of the codebase because their vulnerability‑detection task demands it.

Moreover, in high-efficiency and high-stake situations like smart contract auditing, it often has higher requirements to auditors. Firms like CertiK\footnote{\url{https://www.certik.com/en/products/smart-contract-audit}} often operate under intense time constraints (e.g., 48-hour audits), during which auditors must quickly develop a deep understanding of unfamiliar codebases, in order to identify vulnerabilities. Given the immutability of deployed smart contracts, any missed vulnerability can lead to irreversible financial loss. These pressures have led many auditors to adopt LLMs to accelerate codebase-level comprehension.

In contrast, general developers \cite{bexell2024developers, reefman2024using} often onboard codebases quarterly or monthly \cite{reefman2024using}, work long-term on a smaller set of familiar projects, where code understanding may remain local, partial, or limited to getting the code to run. For example, Reefman et al. \cite{reefman2024using} found that developers used LLMs mainly to get a quick, surface‑level overview or to decode syntax and small snippets. 

Our observation in this difference between code auditors and developers has made us realize that auditors may have tacit knowledge that general developers could learn in performing codebase understanding process.

\subsection{Research Questions}

To answer our key research question described in Section \ref{sec:introduction}, we propose four detailed research questions (RQs) to further guide our study. By learning from code auditors, we are interested in:

\begin{itemize}
    \item RQ1: How do auditors cognitively navigate and understand an unfamiliar codebase? What information are they seeking during the process?
    \item RQ2: What types of assistance do they seek from LLMs (e.g., ChatGPT) during the process? What challenges do they encounter? What suggestions do they have?
\end{itemize}

Moreover, we are also interested in whether the findings derived could yield a set of actionable insights to further inspire the design of future \llmsystem, particularly, we aim to answer: 

\begin{itemize}
\item RQ3: How does our expert-inspired system affect experienced and novice developers’ perceived intuitiveness, ease of use, and others?
\item RQ4: How are experienced and novice developers’ system usage behaviors affected, and why?
\end{itemize}
\section{Methodology Overview}
\label{sec:method}

To answer our questions, as shown in Figure~\ref{fig:method}, we employed a four-stage method from formative interview to prototype development and validation, which is commonly used in HCI research \cite{helander2014handbook, nngroup_formative_summative_2019}. 

\begin{itemize}
    \item \textbf{Stage 1: Formative interview.} To investigate how code auditors understand new codebases, we conducted interviews with eight professional auditors to identify their codebase comprehension strategies. The findings of this stage mainly address RQ1 and RQ2. 
    \item \textbf{Stage 2: Design opportunities}. To support the future development of \llmsystem, we translated our identified professional auditing strategies into four design opportunities through which the LLM system could provide cognitive support in information extraction, decomposition, representation, and user interaction. 
    \item \textbf{Stage 3: Prototype design}. To operationalize our design opportunities, we designed and developed an LLM-powered web application prototype, \systemabb. It incorporates two key features: (1) dynamically \textit{extracting}, \textit{decomposing}, \textit{representing} information aligned with human's hierarchical understanding flow; and (2) allowing users to interact and switch between different abstraction layers, updating the visualization as their understanding evolves.
    \item \textbf{Stage 4: Prototype validation}. To validate the effectiveness of the prototype, we conducted an evaluation with 9 experienced developers and 6 novice developers.  Results from this stage address RQ3 and RQ4, drawing on participants' subjective perceptions captured through post-study questionnaires and verbal feedback elicited during system use.
\end{itemize}


\section{Stage 1: Formative Interview}
\label{sec:formative}


\subsection{Method}
The formative study includes 8 semi-structured interviews with code auditors with a diverse set of auditing experience. All participants had experience with both LLM-assisted and manual code auditing. Demographic information is provided in Table \ref{tab:tab1_interview_participants_demographics}.

\begin{table*}[!htbp]
\centering
\caption{Demographic Information of Interview Participants}
\label{tab:tab1_interview_participants_demographics}
\vspace{-3mm}
\renewcommand{\arraystretch}{1.1}
\resizebox{\linewidth}{!}{\begin{tabular}{@{}cccccccc@{}}
\toprule
\textbf{ID} & \textbf{Occupation} 
& \textbf{Tools Used}
& \textbf{\begin{tabular}[c]{@{}c@{}}Onboarding \\ Frequency\end{tabular}} 
& \textbf{\begin{tabular}[c]{@{}c@{}}No. of Audited\\ Projects\end{tabular}} 
& \textbf{\begin{tabular}[c]{@{}c@{}}LLM \\ Usage\end{tabular}} 
& \textbf{\begin{tabular}[c]{@{}c@{}}Coding \\ Exp. (Years)\end{tabular}} 
& \textbf{\begin{tabular}[c]{@{}c@{}}Auditing \\ Exp. (Years)\end{tabular}} \\ 
\midrule
P1 & Smart Contract Auditor          & Coding Platforms; VSCode Plugin; Auditing Tools & Daily to Weekly   & 200--250 & Daily   & 7   & 3   \\
P2 & Smart Contract Auditor          & VSCode Plugin; Auditing Tools                   & Weekly            & 100--150 & Daily   & 2   & 2   \\
P3 & Engineer                        & VSCode Plugin; Auditing Tools                   & Weekly            & 100--150 & Weekly  & 3   & 1.5 \\
P4 & Smart Contract Auditor          & VSCode Plugin; Auditing Tools                   & Weekly            & 100--150 & Daily   & 2.5 & 2.5 \\
P5 & Smart Contract Auditor          & VSCode Plugin                                   & Weekly            & 50--100  & Daily   & 2   & 2   \\
P6 & Computer Science Master Student & Auditing Tools                                  & Monthly           & 0--50    & Daily   & 3   & 1   \\
P7 & Smart Contract Auditor          & VSCode Plugin; Auditing Tools                   & Weekly to Monthly & 0--50    & Daily   & 4   & 1   \\
P8 & Smart Contract Auditor          & VSCode Plugin                                   & Monthly           & 0--50    & Monthly & 3   & 0.5 \\
\bottomrule
\end{tabular}}
\begin{minipage}{\textwidth}
\footnotesize
Notes: Coding Platforms (e.g., auditwizard.io); Auditing Tools (e.g., slither, echidna, certora)
\end{minipage}
\vspace{-3mm}
\end{table*}

During the interview, we asked participants to reflect upon and share their most notable code auditing practices and experiences of using LLMs to aid code auditing. In addition to narrating their experiences, we requested participants to demonstrate their code auditing process via screen sharing when feasible. We also explored the challenges they encountered with LLMs and their expectations concerning potential AI assistance in the process. All interviews were conducted virtually, audio-recorded, and transcribed for further analysis. 

Following Richards et al.’s collaborative approach to qualitative analysis \cite{richards2018practical}, the interviewer and another qualitative researcher jointly analyzed the data \footnote{Prior work has also explored LLM-assisted approaches to collaborative qualitative analysis \cite{gao2024collabcoder}. We chose a fully manual coding process to ensure analytical rigor and maintain close engagement with the data.}. First, we independently performed open coding on three representative transcripts. Next, we met to compare codes, resolve disagreements, and agree on an initial codebook. We then piloted and refined this codebook on additional data. Finally, one coder applied the finalized codebook to the remaining transcripts. The full codebook and detailed results are available on our \href{https://gaojie058.github.io/code-map/}{website}. We present findings below.

\begin{table*}[htbp]
\centering
\caption{Formative Interview Findings Summary for RQ1 and RQ2}
\renewcommand{\arraystretch}{1.5}
\vspace{-3mm}
\scalebox{0.95}{\begin{tabular}{>{\raggedright\arraybackslash}p{6.5cm} >{\raggedright\arraybackslash}p{5.5cm} >{\raggedright\arraybackslash}p{5.8cm}}
\toprule
\textbf{Understanding Flow and Information Seeking} & \textbf{ChatGPT Assistance ($\checkmark$), Challenges ($\times$) \& User Suggestions ($\star$)} & \textbf{Example Quotes (Slightly modified for readability)} \\
\midrule
\textbf{\yellow{Global Level Understanding.}}
Auditors gather \textbf{project overview} information, including key file names, project background, and auditing requirements, from README files and documentation. \textit{"After obtaining the project, I will read its written content, including white paper and README. I will not read very carefully. I only need to know what each specific contract is for."} P1 & $\checkmark$ Assist in high-level explanation of the codebase; \newline $\checkmark$ Assist in evaluating codebase complexity \newline $\times$ Lack of automated guidance for codebase understanding (requires manual prompting) \newline $\star$ LLM tools should provide pre-defined domain-specific prompts & \textit{"For example, [to assess] how many days we can save and how much money we can charge for this project. Previously, it might have taken two or three days to assess, but now we may only need one or two minutes to get the job done [via LLMs]."} P6 \\
\midrule
\textbf{\yellow{Global Level Understanding.}} Auditors understand \textbf{business logic flow and function-call flow, seeking main components, relationships, key files, functions, and variables.} For example, they examine how files are connected through imports, what inheritance relationships exist, where the entry point is, how data flows through the codebase, and how classes and functions interact. \textit{"You have a general basic [structure] framework... Some companies will draw a diagram, such as architecture diagram."} P6 & $\checkmark$ Assist in understanding which components use a given variable; \newline $\times$ Lack of automated guidance for codebase understanding (requires manual prompting) \newline 
$\times$ Lack of automated information decomposition (requires manual decomposition) \newline $\times$ Lack of filtering for irrelevant code, leading to redundant analysis effort \newline $\star$ LLM tools should provide pre-defined domain prompts  \newline $\star$ LLM tools should provide codebase structure visualization & \textit{"I will go to ChatGPT and ask it which functions use this variable and which functions modify it, so that I can focus on those functions. After understanding some of the functions and roles of global variables, we can then enter each function."} P2  \newline \textit{"Especially when I want to understand the relationships between multiple files or multiple contracts, it probably won't be able to help me. Especially when I input more data to ChatGPT, the response is very broad."} P7\\
\midrule
\textbf{\lightblue{Local Level Understanding.}} Auditors \textbf{locate components, files, variables, and specific contents of interest}, such as important variables and their dependency order, key functions and their call relationships. \textit{"There might be an issue [at a certain point], and I'll make a note saying what kind of issue it is... Then, after finishing looking at this [overall structure], I will come back and focus on this marked part."} P8 & $\checkmark$ Assist in extracting key information from individual files \newline $\times$ Lack of support for cross-file variable relationships \newline $\times$ Lack of automated information decomposition (requires manual decomposition) \newline $\star$ LLM tools should help users identify which parts to focus on & \textit{"At the beginning, I want it to explain to me the specific content of the codebase. I will tell it to explain to me the important state variables, and help me summarize the important functions. I will probably also ask it to summarize certain content".} P7 \\
\midrule
\textbf{\red{Detailed Level Understanding.}} Auditors seek \textbf{clarification of specific code snippets}, verifying whether the code aligns with their current understanding. & $\checkmark$ Assist in explaining code snippets \newline $\checkmark$ Assist in confirming users' understanding \newline $\times$ Lack of continuous understanding flow---clarification questions often interrupt the ongoing reasoning process & \textit{"After giving ChatGPT a code snippet, [I may have] many questions. After you ask question a, when you then ask question b, there is a process interruption. If you ask question b, it will combine the answer to question a to give you the answer to question b."} P5
\\
\bottomrule
\end{tabular}}
\label{tab:formative_study}
\vspace{-3mm}
\end{table*}

\subsection{Results}
\label{sec:formative_findings}

We present our results to answer RQ1 and RQ2 in Table~\ref{tab:formative_study}.

\noindent \textbf{RQ1: How do auditors cognitively navigate and understand an unfamiliar codebase? What information are they seeking during the process?}

\noindent We find that code auditors often follow a hierarchical understanding flow---progressing from global project overview to local modules and detailed implementation checks. 
At the \textbf{global} level, auditors first construct the big picture of the codebase by generating a high-level map of modules, business logic, and dependencies to grasp the overall structure. 
At the \textbf{local} level, they focus on specific modules and key variables that implement critical functionality. 
At the \textbf{detailed} level, they inspect concrete lines of code or execution paths to verify correctness and data flow. 
This gradual reasoning allows them to iteratively refine a coherent mental model of the system and validate their hypotheses across abstraction levels.

\noindent \textbf{RQ2: What types of assistance do they seek from LLMs (e.g., ChatGPT) during the process? What challenges did they encounter? What suggestions do they have?}

\noindent Throughout this process, we find that auditors frequently relied on conversational LLMs such as ChatGPT for summarizing or explaining code fragments. 
However, they reported three recurring challenges: 
(1) the need for repetitive manual prompting to retrieve and understand hierarchical structure, 
(2) the cognitive burden of manually decomposing large projects into promptable segments, and 
(3) irrelevant or overly general answers that failed to align with their reasoning focus. 
As one participant noted, 
``\textit{I can ask the model to explain a function, but it never tells me how this connects to the rest of the codebase.}'' 

Our results highlight that while current LLM tools can explain detailed information, they lack mechanisms to represent structural relationships and support auditors' layered cognitive flow. Code auditors envisioned tools that could automatically organize LLM-generated information according to their cognitive workflow---providing visual overviews, structured decompositions, and explanations that maintain cross-level coherence.


\vspace{-8pt}
\section{Stage 2: Design Opportunities for \llmsystem}
\begin{figure}[!htbp]
    \centering
    \includegraphics[width=\linewidth]{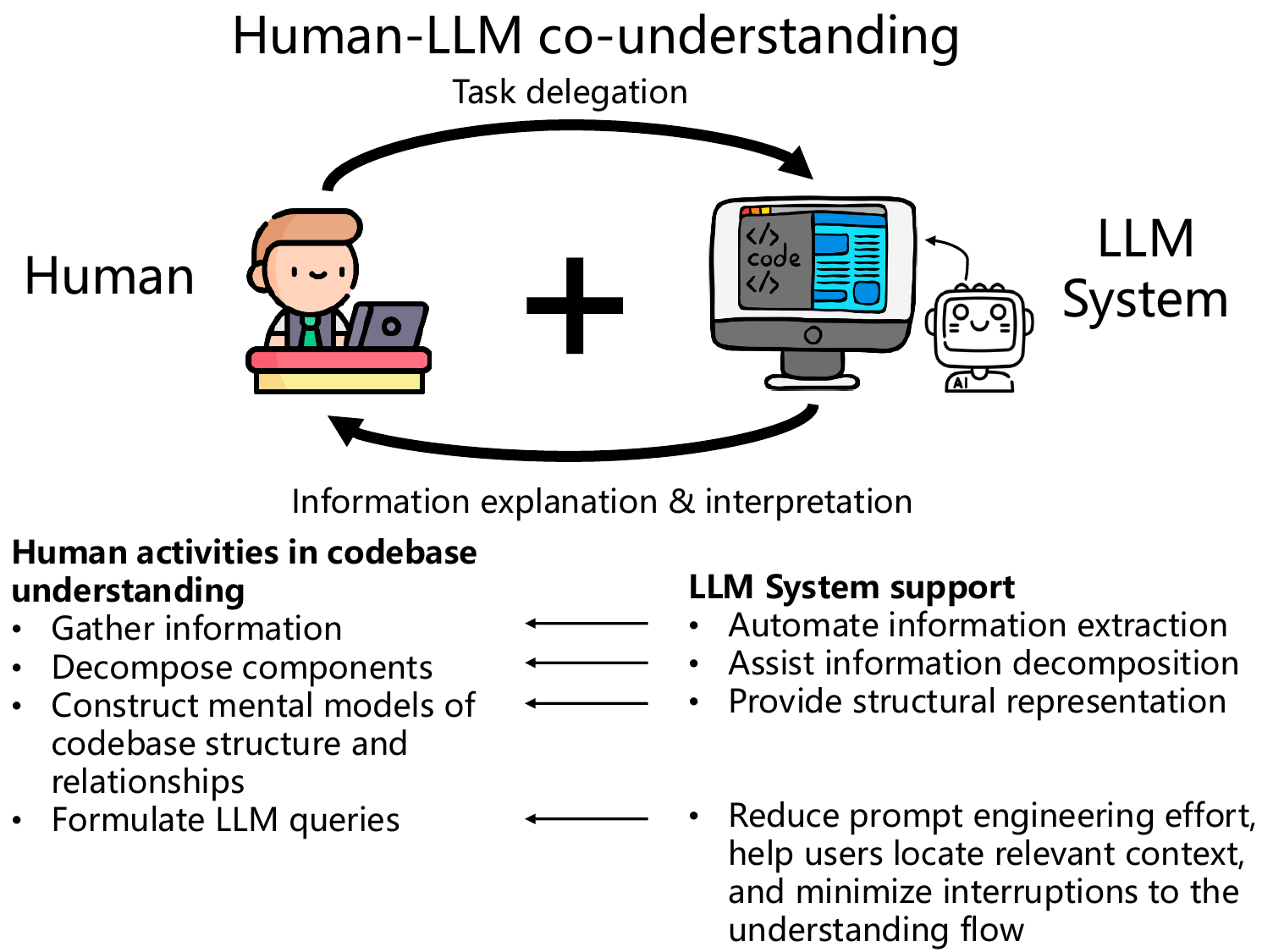}
    \vspace{-20pt}
    \caption{Human-LLM co-understanding of codebases. An LLM-powered codebase understanding system can support human cognitive processes by automating information extraction, decomposition, and structural representation.}
    \label{fig:llm-system-support}
    \vspace{-3mm}
\end{figure}

Since the onboarding process remains challenging, developers may not easily acquire effective codebase understanding skills. Given the practical demands, task complexity, and high learning costs, developers may benefit from an LLM system that provides systematic guidance throughout the understanding process. Below, we summarize the design opportunities derived from our formative study for developing an LLM-powered system to effectively support codebase understanding (also in Figure \ref{fig:llm-system-support}).

\noindent \textbf{Design Opportunity 1:} \llmsystem should \textbf{support information extraction aligned with the human understanding flow.} Instead of directly displaying all information from the codebase, our interviews reveal that users need to manually gather specific information at different levels: global level--project background and structural information; local level--key files and component details; and detailed level--information clarification. This suggests that an \llmsystem should support a structured three-level chain of comprehension, reducing users' manual effort in seeking information across these levels.

\noindent \textbf{Design Opportunity 2}: \llmsystem should \textbf{support information decomposition} to align with the human mental model, deepening understanding within each abstraction layer. Our interview findings show that users need to break the codebase structure into different functional components so they can focus on specific parts for deeper understanding. This process is often manual and requires expertise. Therefore, an \llmsystem should automate information decomposition to reduce the manual effort required from users.

\noindent \textbf{Design Opportunity 3:} \llmsystem should \textbf{support information representation connecting across multiple abstraction layers.} Information in a codebase is inherently complex, spanning multiple files, functional components, and detailed function implementations. Grasping these relationships requires substantial cognitive effort, particularly when reasoning about inter-module dependencies. Therefore, an \llmsystem should provide structural representations across multiple abstraction layers, enabling easier and more efficient comprehension.

\noindent \textbf{Design Opportunity 4:} \llmsystem should \textbf{maintain an analytical structure and scaffold to reduce manual effort in formulating LLM queries and support transitions between the system's structured understanding flow and users' own evolving reasoning.}
Our interview findings show that querying for specific information often distracts users from their original understanding flow when interacting with conversational LLMs. Moreover, users need both expertise and skill to formulate effective queries--for example, locating the specific information they wish to focus on and translating their analytical goals into language that the LLM can interpret. This challenge is more pronounced for novices, who may not possess the same analytical skills or domain expertise as professionals. To address this, an \llmsystem could provide predefined information structures or commonly requested details that are automatically displayed, allowing users to access relevant insights without manually formulating queries.

\section{Stage 3: Prototype Design}
To operationalize the four design opportunities, we designed a prototype, \systemabb. 

\subsection{Usage Scenario}
\begin{figure}[!t]
    \centering
    \includegraphics[width=\linewidth]{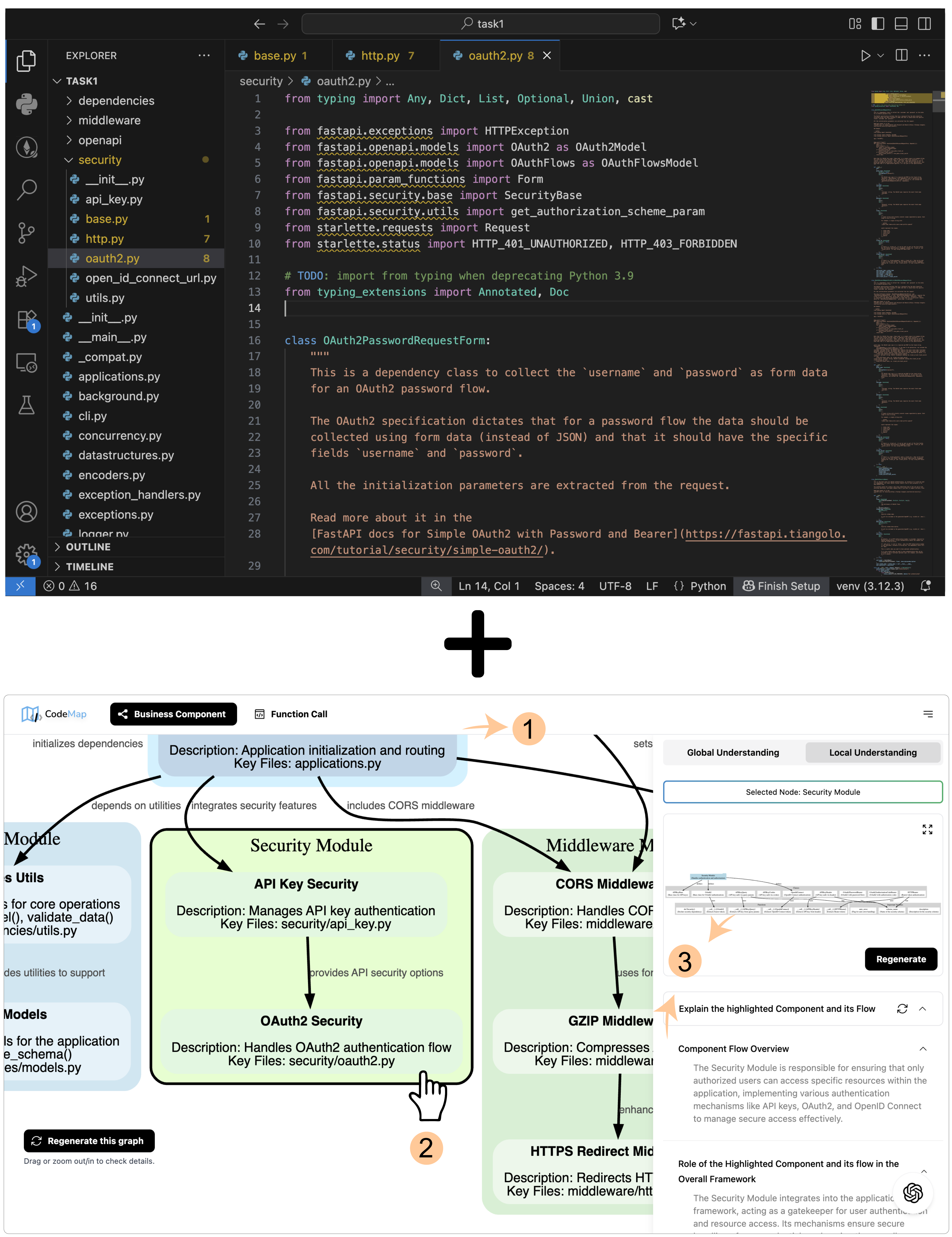}
    \caption{Interface for an example business component map. To support codebase reading, \systemabb allows users to start with 1) Global Understanding through the Global Business Component Map, which provides natural language--enhanced guidance on key modules, components, files, functions, and their relationships. Then, users can 2) zoom into a specific module (e.g., ``Security Module'') to view 3) textual instructions in the right pane that describe the module's details and its relationships with other modules.}
    \label{fig:usage scenario}
\end{figure}

\begin{figure*}[!t]
    \centering
    \includegraphics[width=\linewidth]{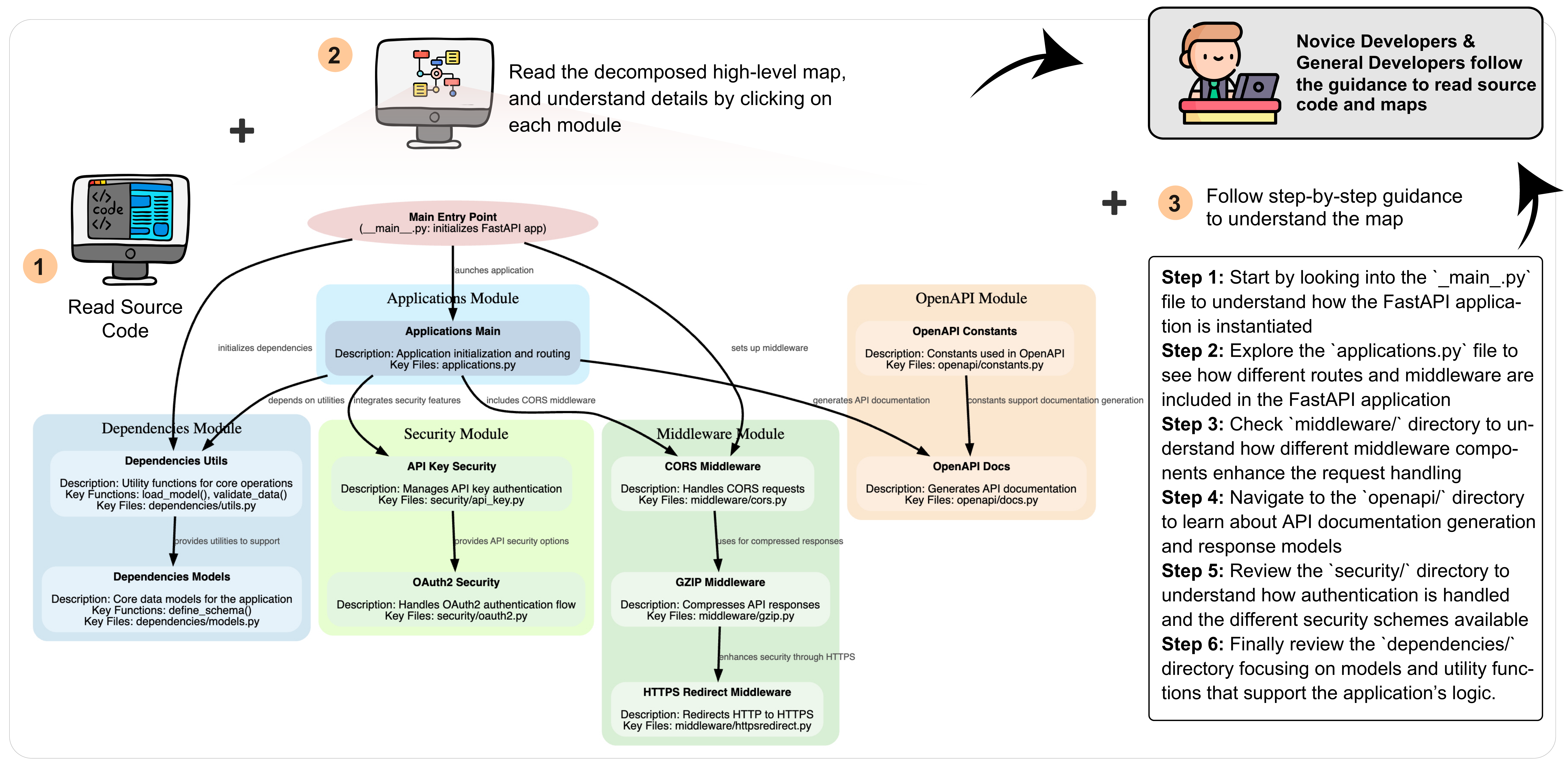}
    \vspace{-6mm}
    \caption{An illustration of how users interact with CodeMap to understand a codebase} 
    \label{fig:usage scenario2}
    \vspace{-10pt}
\end{figure*}

We illustrate CodeMap with a concrete scenario in Figure \ref{fig:usage scenario}, with a demonstration video available on our \href{https://gaojie058.github.io/code-map/}{website}. Suppose a novice developer, \emph{Bob}, needs to understand a new codebase. While reading the source code in VSCode, he uploads the repository to CodeMap. After a brief waiting period, CodeMap generates a high-level business component map of the project. Next, Bob reads the global understanding project overview and follows step-by-step guidance to locate the codebase's entry point and main modules (Figure \ref{fig:usage scenario2}). Meanwhile, he switches back to VSCode to inspect the original code and verify the relationships shown in the map.

Once he has a basic understanding of the global overview, he explores a specific component by clicking its node (e.g., Security in Figure \ref{fig:usage scenario}). A local map is generated (Figure \ref{fig:usage scenario}.3), and the text explanation summarizes the module's functions and relationships with other modules.
Additionally, this focused view situates the module within the broader business flow, allowing him to inspect details without losing context.

If Bob wants to delve deeper into a specific detail, for example, the purpose of a local variable, he can click the variable and ask the chatbot (powered by an LLM) a question such as, \textit{“Hi CodeMap, can you tell me more about this node?”} CodeMap then returns an answer focused on that particular node.

Similarly, for a deeper analysis of the codebase, Bob can open the function-call view to examine relationships between classes, following the same interaction pattern described above.

\subsection{Key Features}

\noindent \textbf{Key Feature 1: Extracting and representing information at each layer.}
\systemabb extracts the desired information and represents it in formats tailored to what users need at each layer:
\begin{itemize}
    \item \textit{Global map}: \textit{business component map}: extracting high-level structural information, including business modules, components, and links to show their relationships; \textit{function call map}: visualizing classes as nodes and showing their inheritance relationships.
    \item \textit{Local map}: helping users to understand the relationships between variables and functions within each file.
    \item \textit{Natural language enhancement for readability}: presenting the links between nodes in the map and natural language to describe the relationships between modules. For example, \textit{“Component A provides or manages data for Component B”}. Moreover, providing step-by-step guidance for global map to orient users through project's modules.
\end{itemize}
In this way, users no longer need to gather information manually and can focus their efforts on understanding the overall structure of the codebase.

\noindent \textbf{Key Feature 2: Seamless switching between layers through interaction.}
\begin{itemize}
    \item \textit{Switch between layers through user interaction.} Once the three-layer understanding structure is constructed, \systemabb allows users to easily switch between the global, local, and detailed query levels, or move from detailed queries back to the local or global views. This switching is enabled through simple clicks, unlike traditional conversational LLM interactions, which are typically linear and make it difficult to return to higher-level contexts after a query. 
    \item \textit{In-context Query:} When the user submits a conversational query at any layer, \systemabb gathers relevant context, such as module names, files, to formulates a response based on the user's selected code area. This enables users to interact naturally while preserving structural context. In this way, users can explore deeper at each layer while maintaining a coherent understanding flow.
\end{itemize}

\subsection{Implementation}
\begin{figure}[!hb]
    \centering
    \includegraphics[width=\linewidth]{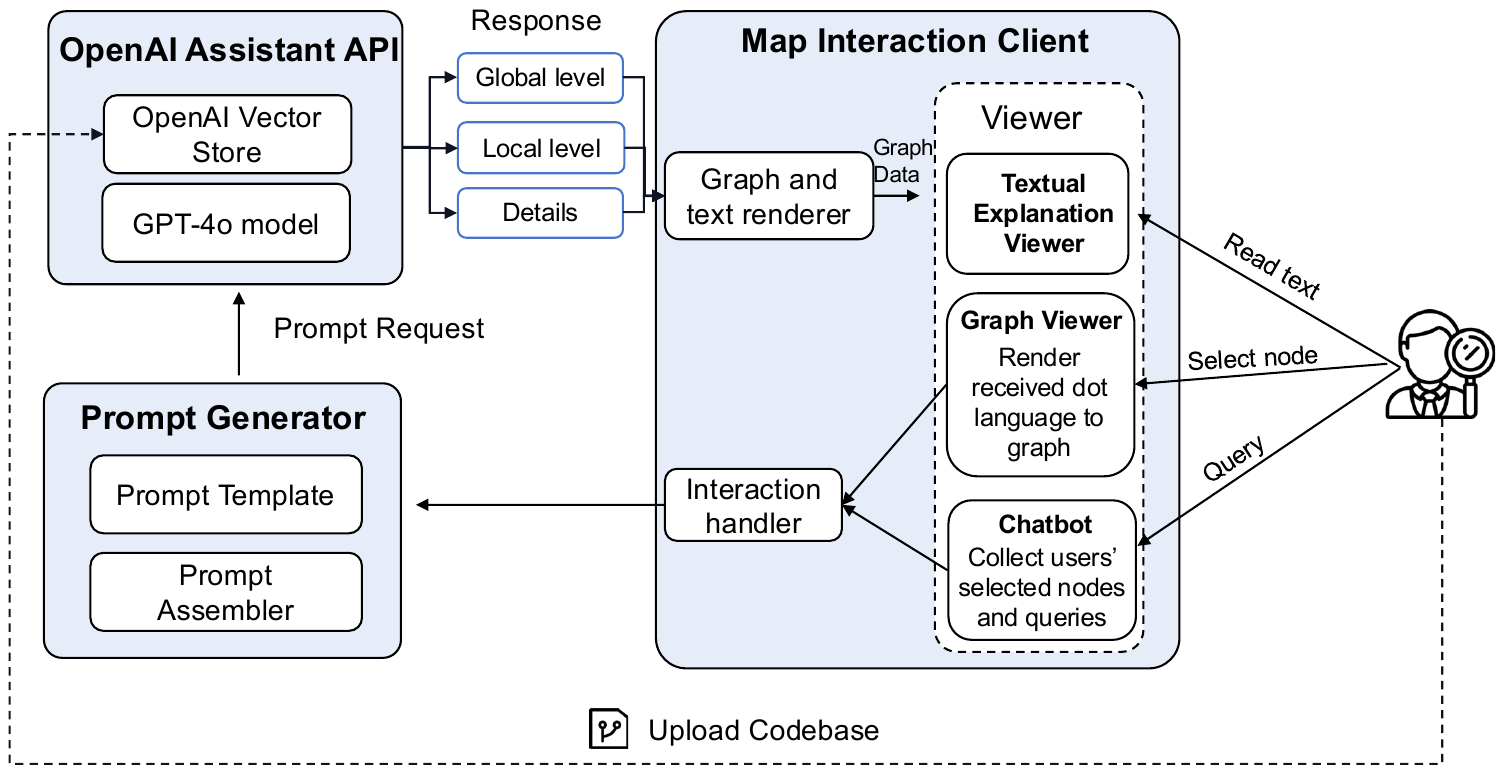}
    \vspace{-8mm}
    \caption{CodeMap implementation overview}
    \label{fig:system-structure}
    \vspace{-10pt}
\end{figure}

\textbf{Web Application.} We implemented \systemabb as a web application to facilitate easier testing later. Its system architecture is illustrated in Figure \ref{fig:system-structure}. It includes 1) the prompt-generation backend: including prompt templates, prompt assembler, and retrieval-augmented generation (RAG) system; 2) the map-interaction client.
First, the backend assembles structured  prompts from the user’s interaction and prompt template. These prompts are then sent to RAG system, which employs vector store and LLMs to retrieve vectorized source-code chunks and produce context-aware answers. Specifically, we chose GPT-4o model for LLM due to generation efficiency, which provides native support for RAG via the OpenAI Assistants API \cite{openai2025assistanttools}.
\systemabb first uploads the entire codebase to the Assistant API with \texttt{purpose=assistants} so it is indexed for retrieval, then issue queries via the \texttt{thread/message} endpoints.
On the frontend, \systemabb is implemented with the React framework. For real-time rendering of both the call graph and the accompanying textual explanations, it used D3-Graphviz \cite{jacobsson2020d3graphviz}. Finally, users can view and interact with the graph through node selection, querying, and graph regeneration. In particular, once a node is selected, \systemabb collects the node’s details and submits them to the RAG system, which then returns local understanding information.

\begin{figure}[!hb]
    \centering
    \includegraphics[width=\linewidth]{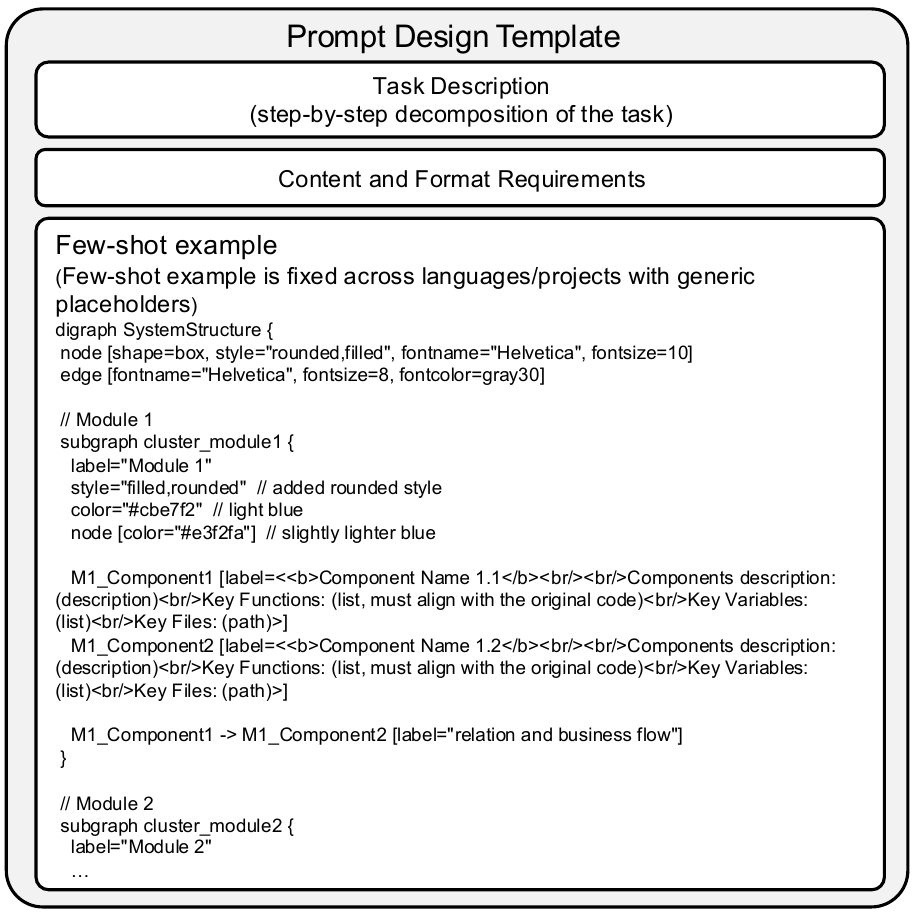}
     \vspace{-6mm}
    \caption{Prompt structure for global business understanding}
    \label{fig:prompt-design}
    \vspace{-10pt}
\end{figure}

\begin{figure*}[!htbp]
    \centering
    \includegraphics[width=\linewidth]{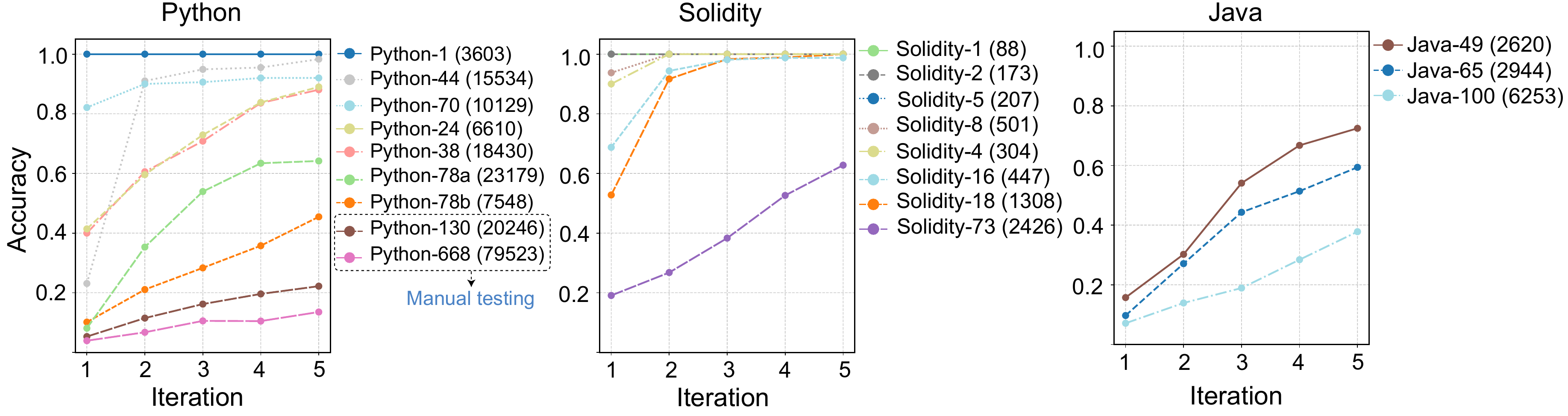}
    \caption{Accuracy Evaluation Results. The legend for each line indicates the programming language along with the number of files, and the total lines of code within brackets. For example, Python-44 (15534) represents a Python project with 44 files and 15,534 lines of code.}
    \label{fig:accuracy}
     \vspace{-5mm}
\end{figure*}

\noindent \textbf{Prompt Design.} We carefully designed prompts to ensure the generation of both maps and textual explanations. Taking global understanding as an example (Figure \ref{fig:prompt-design}), it involves generating a business map (DOT) \cite{graphviz2025dotlang} and an overview (JSON). To ensure proper generation of both DOT and JSON, the prompt includes a task description, explicit output constraints, and a few-shot example to guarantee strict adherence to the desired format. In the JSON output, it provides a summary, entry-point identification, execution instructions, and a step-by-step architectural reading guide that helps users navigate the global understanding logically.

Notably, our prompts are neither project-specific nor domain-specific. They rely on few-shot examples in both JSON and DOT formats with generic placeholders (e.g., \texttt{Module1}→\texttt{Module2}) and deliberately omit details tied to any specific programming language or codebase. Consequently, the same prompts can be applied to both domain-specific languages (e.g., Solidity) and general-purpose languages (e.g., Python).

Furthermore, prompts for additional features, such as the global function-call map and local function-call map, follow a similar structure and can be found on our project \href{https://gaojie058.github.io/code-map/}{website}.

\subsection{Accuracy Reliability Checking}
While convenient, relying entirely on LLMs for content generation is unreliable because of LLMs’ limited RAG capabilities (e.g., missing components or files) and their tendency to hallucinate. We designed an accuracy ensuring mechanism to alleviate this limitation.

\noindent \textbf{Iterative Refinement.} We implemented an iterative refinement process--commonly used for LLM-generated outputs \cite{gu2024airesilient, liu2024howai}—for graph generation; each iteration builds upon the previous iteration's output. The iteration prompts are available on our \href{https://gaojie058.github.io/code-map/}{website}.

\noindent \textbf{Accuracy Evaluation.} There is no definitive ground truth for the business component map, as different people may have varying interpretations and decompositions of its functionalities. Therefore, we decided to use file counts within the codebase, which are fixed, as one form of ground truth--meaning that if the generated map contains no incorrect files and fewer missing files, it can be considered more accurate. We compare the files contained in each LLM-generated graph with the file ground truth to approximate the overall accuracy of the map\footnote{Accuracy is calculated as $\text{Accuracy} = \frac{\text{TP} + \text{TN}}{\text{TP} + \text{TN} + \text{FP} + \text{FN}}$ \cite{baratloo2015part}, where TP represents the number of correctly retrieved files. In our context, only TP, FP, and FN are meaningful. TN is not applicable since no files in the codebase are expected to be excluded from the map. Ideally, all files uploaded to the vector store will be analyzed and represented in the graph. The closer the number of files in the graph generated by \systemabb approximates the ground-truth number of files in the codebase, the higher the accuracy should be. }.

\noindent \textbf{Evaluation Setup.} To evaluate \systemabb on real projects with diverse codebase sizes and programming languages, we selected 9 Python, 8 Solidity, and 3 Java open source projects. 
For each project, we generated maps for 5 refinement rounds $\times$ 10 runs independently. We then computed the accuracy averaged over these 10 runs. The formal experiment required more than 1000 Assistant-API calls (excluding preliminary trials). Notably, because OpenAI's vector store limits each upload to only 100 files, we manually evaluated \textit{Python-130} and \textit{Python-668} through ChatGPT user interface.

\noindent \textbf{Results.} Results are shown in Figure \ref{fig:accuracy}. Overall, additional iterations incrementally improve accuracy. Accuracy stabilizes by the fourth iteration for Python and by the third for Solidity, whereas it has not yet stabilized for the tested Java projects. Moreover, this stabilization trend seems to correlate with codebase size, particularly the number of files: smaller codebases--especially Solidity projects--reach high accuracy after fewer iterations, while larger ones (e.g., \textit{Python-668}, \textit{Solidity-73} and \textit{Java-100}) require more rounds and exhibit slower gains.

\noindent \textbf{Best Performance Testing.} Since \systemabb's prompts still struggle with larger projects, we wonder whether forthcoming models with longer context windows might alleviate this problem. We then simulated such models by running \systemabb with GPT-4.1\footnote{One of the most capable models available when we conducted this experiment in 2025}, which offers a 1 million-token context window \cite{openai2025gpt4dot1}. We implemented the retrieval pipeline in LangChain because it supports advanced file chunking and filtering\footnote{We used OpenAI's native vector store for file search in previous evaluation because it requires fewer parameters and ensures experimental reproducibility.}. Our results showed that through five iterative refinement rounds, \systemabb's map generation can achieve 99\% accuracy on \textit{Python-130} and 27\% on \textit{Python-668}. These findings imply that larger context windows can substantially boost \systemabb's accuracy; thus, current limitations may be alleviated as LLM's capability continues to grow.

\section{Stage 4: Prototype Validation}
To validate the feasibility of \systemabb in real developer settings, we conducted a small-scale observational study with 9 experienced developers and 6 novice developers. The goal was to observe how \systemabb influenced their comprehension process and tool feature usage preferences, rather than to conduct a formal performance comparison. Each participant received a reward equivalent to USD 10-15 upon completion of the study.

\subsection{Method}
\subsubsection{Participants}
\noindent \textbf{Experienced developer participants (EP).} We recruited experienced developer participants from our university and public social media platforms. They included PhD and Master's students specializing in general software engineering (EP1, EP2), large language models for software engineering (EP3), and machine learning (EP4, EP5), as well as professionals in bug mining and vulnerability repair (EP6-EP9). Participants reported 4–10 years of practical programming experience in industry or research projects: EP7-EP9 (6 years), EP6 (7 years), EP4 (8 years), EP2 (4 years), and EP1, EP3, and EP5 (10 years). All participants used ChatGPT daily and primarily programmed in Python, C, C++, and Java. \textbf{Novice developer participants (NP).} For novice developers, we recruited 6 participants (4 undergraduate students and 2 master's students) from our university's internal platform to join an in-person study. Their majors included project management (NP1), quantitative finance (NP2), engineering (NP3), data science (NP4), and computer science (NP5, NP6). Participants reported one to five years of programming learning experience (without professional development experience). Most were familiar with Python (NP1-NP4), while some also used Java (NP5) or JavaScript (NP5, NP6). Their ChatGPT usage ranged from monthly (NP3) to weekly (NP1, NP2, NP4) and daily (NP5, NP6).

\subsubsection{Procedure}
\textbf{Experienced participants:} Each participant performed short codebase understanding tasks using VSCode (each lasting 25 minutes), and with the assistance of \systemabb and two baselines: a conversational LLM (i.e., ChatGPT) and a commercial static visualization tool (\textit{Understand}). When using each tool for codebase understanding, participants were asked to think aloud and to complete short comprehension quizzes and NASA-TLX subjective load questionnaire as \textit{reflective probes} to prompt purposeful interaction and articulation of understanding and usage experience. Each session lasted approximately 90 minutes, and all interactions were screen recorded for later analysis.  \textbf{Novice participants:} For novices, given their limited programming expertise, we did not require them to complete the quiz or use the static analysis tool. Instead, they were asked to explore \systemabb and ChatGPT freely in 10 minutes using an example project and to provide feedback.

\subsubsection{Data analysis}
To analyse the user' behaviors, we primarily rely on participants’ experiences through their qualitative verbal feedback, post-study preference questionnaire. For video recordings, we conducted a thematic analysis \cite{maguire2017doing} of the transcribed data. Two authors collaboratively performed the analysis, engaging in several rounds of team discussions to reach a consensus on interpretation. Moreover, to understand how the system influences expert users' behavior, we manually analyzed expert users' feature usage time and how their behavior changed within the 25-minute fixed time window for each task. We manually annotated the time spent on each activity through video analysis. In total, we spent 25 hours on video annotation, analyzing 18.48 hours videos.

\subsection{Results}

\subsubsection{\textbf{RQ3: How does our expert-inspired system affect experienced and novice developers' perceived usefulness, ease of use, and others?\\[-6pt]}}

\begin{figure*}[!t]
    \centering
    \includegraphics[width=\linewidth]{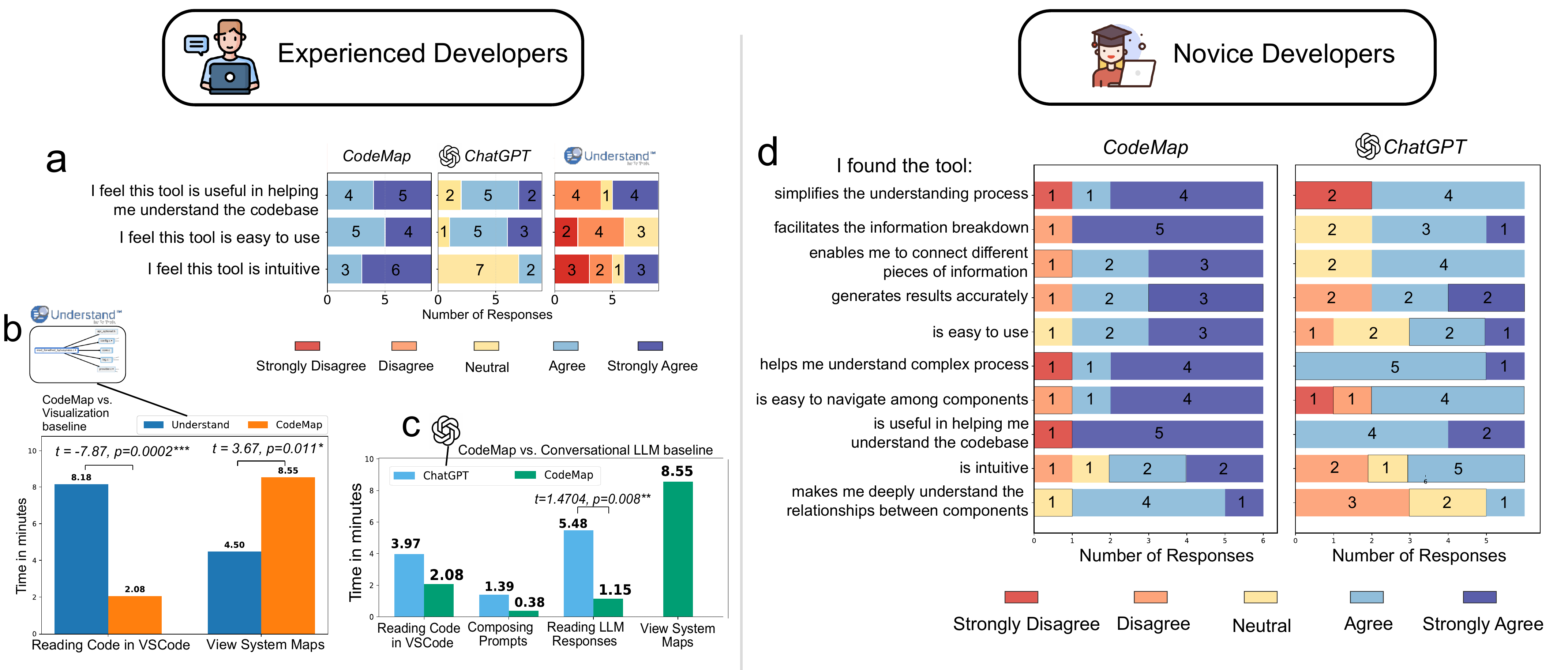}
    \vspace{-4mm}
   \caption{Results: (a) experienced developers’ usability ratings (due to time constraints within the 90-minute task to avoid fatigue, they only rated usability); (b) and (c) feature usage time; (d) novice users’ subjective preference.}
    \label{fig:feedback}
     \vspace{-3mm}
\end{figure*}

\noindent \\ \textbf{Experienced Developers}. As shown in Figure \ref{fig:feedback}a, all experienced participants (9/9) agreed that \systemabb was useful for code understanding, compared with 7/9 for ChatGPT and only 4/9 for \textit{Understand}. All of them (9/9) perceived \systemabb as more intuitive than ChatGPT (2/9) and \textit{Understand} (3/9). Ease of use ratings were similar between \systemabb (9/9) and ChatGPT (8/9), and considerably higher than \textit{Understand} (0/9). 

\noindent \textbf{Novice Developers.} Similarly, novice participants rated \systemabb higher across all questions. 4/6 found the tool intuitive, and 5/6 reported it was easy to use. Most novices noted that \systemabb helped them break down information, simplify the code comprehension process, useful in helping understand the codebase, etc. However, one participant (NP2), who was from quantitative finance background and had never been exposed to full codebase reading, found \systemabb overwhelming due to the amount of displayed information. She preferred ChatGPT, describing its explanations as clearer and its usability as better than our query feature. This suggests that while \systemabb is promising, it may present a learning curve for users with limited programming experience.

\subsubsection{\textbf{RQ4: How are experienced and novice developers' system usage behaviors affected, and why?\\[-6pt]}}
\noindent \\ \textbf{Key Finding 1: A shift from text-based to map-based usage occurs when using \systemabb but does not occur when using the static visualization tool.} We performed a Shapiro–Wilk test on the collected time data and confirmed that all measures followed a normal distribution ($p > .05$). 

As shown in Figure \ref{fig:feedback}b, compared with the visualization baseline (\textit{Understand}), participants using \systemabb spent 75\% less time reading code in VSCode (2.08 min vs. 8.18 min; $t = -7.87$, $p = .0002$) and 90\% more time viewing system-provided maps (8.55 min vs. 4.50 min; $t = 3.67$, $p = .011$), suggesting that \systemabb's natural language-driven visual representations effectively encouraged users to rely more on structured visual information. 

As shown in Figure \ref{fig:feedback}c, compared with the conversational LLM baseline (\textit{ChatGPT}) (5.48 min; $t = 1.47$, $p = .008$), participants using \systemabb spent significantly less time reading LLM responses (1.15 min), representing a 79\% reduction. Instead, they relied more on viewing system maps to build understanding.

In particular, our qualitative feedback showed that, traditional static analysis tools like \textit{Understand} often have interfaces that do not align with the natural flow of human codebase reading and fail to present information at an appropriate cognitive level, which can leave users feeling disoriented, ultimately make them give up and switch to VSCode for direct manual code reading. Like EP7 mentioned, \textit{"For \textit{Understand}, my initial impression is that it has a lot of features...The interface feels complex and cluttered. I’d prefer to use call graph generation plugin and viewing maps in VSCode."}

\noindent \textbf{Key Finding 2: The shift from text-based to map-based usage occurs because \systemabb can 1) decompose and 2) use natural language to explain modules and their relationships.} The high-level map with modules allows users to view the codebase at the module level via semantic descriptions.  
EP7 described, \textit{"It allows me to intuitively understand the entire project by dividing into modules and shows what each module does. It provides very semantic descriptions, which make it very clear."} 

The decomposition not only helps build local understanding but also helps users define a specific context--a challenge to define and describe when using ChatGPT.
As EP8 mentioned, \textit{"We usually just upload the entire repo to GPT and let it analyze everything...it definitely loses focus. If we could instead [use \systemabb to] limit its context by selecting specific modules, I think that's really great."}

In contrast, the information in traditional static analysis tool was not presented appropriately--it lacked key points and textual explanations (EP4, EP5, EP6). The amount of information was overwhelming and difficult to process (EP9), and the call relationships lacked appropriate granularity, making them too complex to understand (EP3). \textit{Understand} could be suitable for development scenarios where users already built a high-level understanding of the codebase and are familiar with files and module relationships (EP8).

\noindent \textbf{Key Finding 3: The shift from text-based to map-based usage may occur because \systemabb requires less expertise and effort for prompting and may offer greater usefulness in the long term.} ChatGPT may only gradually build understanding with already established expertise (EP3). This includes the ability to ask appropriate questions based on the codebase (EP2, EP3, EP7), knowing what they need to know (EP3, EP7, EP8), and using precise questions, including terms from the codebase, to compose their prompts with prompt engineering skills. Like EP3 mentioned, \textit{"ChatGPT just analyzes for me whatever I tell it to analyze. So I need to know what I want to analyze."}
Without expertise and a clear purpose, the textual explanation can sometimes just list categories without deep interpretation (EP2) or generate answers based on surface-level information, such as file names, rather than deeply understanding the entire content (EP6). 


\noindent \textbf{Key Finding 4: \systemabb may demonstrate better effectiveness when applied to large-scale and more complex projects} (EP2, EP3, EP4, EP6, EP7). Although our validation study involved relatively small tasks, participants noted that \systemabb would likely be more useful for larger projects. For example, projects with over 10,000 lines of code where ChatGPT often fails to illustrate the relationships between modules (EP7). As EP4 mentioned, \textit{"The reason you’d create diagrams is when a project is really big — when it’s too hard to describe clearly using only language."} Moreover, when projects become larger and more complex, it is difficult for humans to draw high-level maps and enrich with details. As EP3 explained, \textit{"If there are a lot of modules, I might roughly draw a diagram [with a few blocks] myself, but I wouldn’t go as deep as your second layer [local understanding]."}

\noindent \textbf{Key Finding 5: \systemabb's understanding flow can decrease novices' non-purposeful reading and help them build understanding progressively.}
For example, NP4, an undergraduate student with some basic coding knowledge, felt lost when interacting with ChatGPT and asked questions like \textit{"Can you explain starlette?"}\footnote{Starlette is a lightweight ASGI (Asynchronous Server Gateway Interface) framework for building high-performance web applications and APIs in Python.}. Instead, 
in this lost situation, \systemabb can provide participants a "starting point". Like NP5 said, \textit{"This [project architecture understanding guide] is so amazing because when you handle a project for the first time and have no idea what it is about, this is a very good guide to help you understand it."}

Maps are easier for them to learn. NP6 explained, \textit{"If you want to learn something new and difficult, it is easier to learn through a diagrammatic approach, and it is easier to understand using a diagrammatic approach like a UML class diagram, which ChatGPT fails to give."}
Similarly, NP1 mentioned the maps are aligned with his experiences in homework assignments, NP1 will draw some simple maps with a few classes.
Moreover, \systemabb also provides them with a mental model they can follow, helping ensure they do not miss important parts. As NP1 said, \textit{"If you just read the code by yourself, you might miss some details like some key functions. Also, because it's a very large project, like a game project with hundreds of lines of code, right? So this global understanding can provide me with a blueprint-like view and help me understand the structure."}

\noindent \textbf{Key Finding 6: \systemabb can help novices avoid superficial results from interacting with ChatGPT}. Aligning with our formal user study, many novices, when interacting with ChatGPT for analysis, tend to ask simple, straightforward questions, such as the meanings of specific terms (NP4), without understanding it from a big picture. NP6 also mentioned that ChatGPT requires users to actively output and know what to ask. 
Like he mentioned, \textit{"It will ask, `What do you want?' But in \systemabb, a set of standard questions is already present, such as `What is the business problem about?', `What are the relationships between components?', and `How does the flow work?' Everything is already directly presented in \systemabb. If we want to ask more questions, there is the GPT button — you can type and ask about it."}
Moreover, with long descriptions from ChatGPT, participants mentioned that the responses were overly descriptive, and they often did not know which parts were correct (NP6).

\section{Discussion}

\subsection{Human-AI Co-understanding for Codebase}

We conceptualize \systemabb as facilitating human-AI co-understanding of codebases (Figure \ref{fig:llm-system-support}), where AI acts not merely as an answer generator but as a collaborator that extracts, reformulates, and contextualizes information in ways aligned with human comprehension processes. Through this collaboration, machine-level representations can be converted into human-readable structures, while humans iteratively steer the exploration through interpretation and inquiry. This loop turns comprehension into a shared activity rather than a one-way automation. \systemabb's design shifts the interaction paradigm from standard prompting to UI-enhanced and context-based modes, as categorized in prior taxonomies of human-LLM interaction \cite{gao2024taxonomy}. 

\vspace{-8pt}
\subsection{Generalizability and Use Cases} 
The understanding flow we identified is inherently broader in scope, mirroring general activities such as reading a book, grasping complex concepts. This aligns with established theories in UX design, like the visual hierarchy principle \cite{still2018web}, and cognitive science theories, such as information-foraging theory \cite{InformationForagingNNGroup}. We envision that in different scenarios, \systemabb can support team task delegation and evaluation by assessing project complexity. In educational contexts, since \systemabb incorporates professional expertise in codebase understanding, it can also be applied to a variety of code-related learning tasks. For example, students working with unfamiliar codebases can receive more effective guidance and feedback by interacting with \systemabb.


\section{Threats to Validity}


\textbf{Internal validity.}
\systemabb’s performance depends on the retrieval accuracy of its RAG component and the capabilities of the underlying LLM, including reasoning ability, context length, and susceptibility to hallucination. We mitigated these factors through iterative refinement and validated our approach using GPT-4.1, which demonstrated clear gains with improved model capabilities. Moreover, we involved 8 auditors for interviews, 15 participants (9 experienced, 6 novice) for validation, comparable to similar studies (e.g., 10 \cite{brachman2025towards}, 12 \cite{xie2024waitgpt}), but the sample size still limits representativeness. Larger and longer-term studies could further strengthen generalizability.
\textbf{Conclusion validity.} We compared \systemabb only with ChatGPT and \textit{Understand}. The effectiveness of other LLM-based tools, such as AI-powered IDEs (e.g., Cursor \cite{Cursor}), remains unexplored. However, current AI IDEs mainly assist with context selection and lack \systemabb's hierarchical visualization and cognition support, limiting their effectiveness for deep code comprehension.

\section{Related Work}
\noindent \textbf{Codebase Understanding.} Codebase understanding is a foundational yet challenging task in software development. It often arises when a developer is newly onboarded to a codebase and needs to become familiar with it in order to identify vulnerabilities \cite{xiang2015overview}, make contributions \cite{reefman2024using, hines2023codebase}, or just learn the code \cite{busjahn2013use}. 
Unlike snippet-level code comprehension, codebase understanding emphasizes the architecture, components and their relationships \cite{levy2021understanding}. Traditionally, codebase understanding is aided by static or dynamic analysis tools that generate structural visualizations (e.g., data flows \cite{yin2002program, scitools2025understand, codecompass2018framework, sourcetrail2021explorer, hines2023codebase} and trace graphs \cite{cornelissen2010controlled}). However, these tools are inherently limited by the incompleteness of program analysis and often incur high processing overheads \cite{besset2018sourcetrailue4}. Users also often report steep learning curves and usability challenges \cite{reefman2024using}.

\noindent \textbf{LLM-powered Codebase Understanding.} Recently, LLM-assisted snippet-level code understanding has attracted significant attention, including efforts to explain selected code snippets \cite{andryushchenko2024leveraginglargelanguagemodels,theprogrammersassistant,chen2023gptutorchatgptpoweredprogrammingtool,nam2024usingllmhelpcode,richards2024you} and related concepts \cite{nam2024usingllmhelpcode,andryushchenko2024leveraginglargelanguagemodels}. 
However, limited work has \cite{reefman2024using, berhanu2025ai} explored the application of LLMs for codebase understanding, which is essential in many software engineering scenarios. Reefman et al. \cite{reefman2024using} incorporate the LLM with static‑analysis context (call hierarchies, repository structure, etc.) to answer repository‑scale developer queries.
However, because the system is implemented purely as a chatbot style interface, it lacks designs for human usage in real world.
To address this gap, we adopted a human-centered perspective by examining how code auditors--who frequently onboard to new codebases--develop code comprehension strategies and how these insights can inform system design.

\section{Conclusion}
This study adopted a human-centered approach to explore how professional code auditors comprehend codebases and how their strategies can inform \llmsystem design. We proposed four design opportunities for LLM-based systems and operationalized them through \systemabb, a prototype that extracts, decomposes, and represents codebase information to support hierarchical understanding. User validation with experienced and novice developers showed that \systemabb enhances intuitiveness, ease of use, and usefulness, while reducing reliance on textual LLM responses. These findings demonstrate the potential of expert-informed, human–AI co-understanding systems to improve large-scale codebase comprehension and developer onboarding.

\section{Data Availability}
\label{sec:data_availability}
All data and code are available on our \href{https://gaojie058.github.io/code-map/}{website}. This study has been approved by the Institutional Review Board (IRB) at the authors’ institution.



\bibliographystyle{ACM-Reference-Format}
\bibliography{reference}


\end{document}